# Time-domain decoding of unconventional charge order mechanisms in nonmagnetic and magnetic kagome metals

Seongyong Lee[1,2], Byungjune Lee[1,2], Hoyoung Jang[3], Xueliang Wu[4], Jimin Kim[1], Gyeongbo Kang[3], Choongjae Won[1,5], Hyeongi Choi[3], Sang-Youn Park[6], Kyle M. Shen[7,8], Federico Cilento[9], Aifeng Wang[4], Jae-Hoon Park[1,2†] & Mingu Kang[1,7,8,10†]

[1]Max Planck POSTECH/Korea Research Initiative, Center for Complex Phase of Materials, Pohang 37673, Republic of Korea.
[2]Department of Physics, Pohang University of Science and Technology, Pohang 37673, Republic of Korea.
[3]X-ray Free Electron Laser Beamline Division, Pohang Accelerator Laboratory, Pohang University of Science and Technology, Pohang 37673, Republic of Korea.
[4]College of Physics and Center of Quantum Materials and Devices, Chongqing University, Chongqing 401331, China.
[5]Laboratory for Pohang Emergent Materials, Pohang University of Science and Technology, Pohang 37673, Republic of Korea.
[6]Pohang Accelerator Laboratory, Pohang University of Science and Technology, Pohang 37673, Republic of Korea.
[7]Laboratory of Atomic and Solid State Physics, Department of Physics, Cornell University, Ithaca, New York 14853, USA.
[8]Kavli Institute at Cornell for Nanoscale Science, Cornell University, Ithaca, New York 14853, USA.
[9]Elettra–Sincrotrone Trieste S.C.p.A., S. S. 14, km 163.5 in AREA Science Park, Trieste 34149, Italy.
[10]Department of Physics and Astronomy, Seoul National University, Seoul 08826, Republic of Korea.

†Corresponding author. Email: iordia@snu.ac.kr, jhp@postech.ac.kr

**In kagome lattice materials, quantum interplay between charge, spin, orbital, and lattice degrees of freedom gives rise to a remarkably rich set of emergent phenomena, ranging from unconventional charge order and superconductivity to topological magnetism. While the exact nature of these exotic orders is often challenging to comprehend in static experiments, time-resolved techniques can offer critical insights by disentangling coupled degrees of freedom on the time-axis. In this work, we demonstrate that the nature of charge orders in two representative kagome metals – nonmagnetic $ScV_6Sn_6$ and magnetic FeGe – which has been highly controversial in static studies, can be directly deciphered in the time-domain through their fundamentally distinct order parameter dynamics measured via time-resolved X-ray scattering at an X-ray free electron laser. In nonmagnetic $ScV_6Sn_6$, the dynamics are characterized by ultrafast melting and coherent amplitudon oscillations, typical of a phonon-coupled charge order. In stark contrast, magnetic FeGe exhibits resilient metastable charge order dynamics, hitherto unobserved in any other charge-ordered system – this unique time-domain behavior directly signifies an unconventional magnetism-interlocked charge order state realized in this kagome magnet. Our results not only provide a model case where unconventional nature of electronic order, hidden in equilibrium, is directly unraveled in the time-domain, but also pave the way for future out-of-equilibrium engineering of novel quantum orders in kagome lattice platforms.**

Charge order is one of the most prevalent emergent quantum phases in condensed matter systems, manifesting in unconventional superconductors,[1] transition metal di-, tri-, and tetra-chalcogenides,[2] twisted Moiré superlattices,[3] quasi-one-dimensional systems,[4] and elemental metals.[5] Despite this prevalence, however, understanding the exact origin of charge order in quantum materials remains a highly nontrivial task, except in a few toy model cases described by Peierls' theorem.[6] This challenge is fundamentally connected to the inherent multicomponent nature of charge order, where electron organization and lattice distortion always appear concomitantly due to electron-phonon coupling, making disentangling their respective contribution challenging. In systems where spin degrees of freedom are also present, such as in elemental Cr or Nickelates,[5,7] spin-charge and spin-phonon coupling also contribute to the ordering, making the physics even richer yet complicated.

Over the past few years, charge order phase has been actively discovered in various nonmagnetic and magnetic kagome lattice systems[8–10] – $AV_3Sb_5$ ($A$ = K, Rb, Cs), $ScV_6Sn_6$, and FeGe – offering a fresh opportunity to study the mechanisms of charge order formation and its interaction with other neighboring phases such as superconductivity,[11] magnetism,[12] and pseudogap.[13] A central yet unresolved question in this field is whether the nature of charge order is *universal* or *distinct* across different kagome systems. On the one hand, the charge order can emerge from a generic electronic instability associated to the van Hove singularities (vHS) in the kagome electronic structure.[14] This scenario is particularly interesting as this vHS-driven charge order is subject to sublattice interference mechanism,[14,15] and can accompany more exotic phenomena, such as time-reversal symmetry breaking orbital-current order and many-body-induced nontrivial topology. Experimentally, vHSs at the Fermi level[12,16,17] as well as evidence of time-reversal symmetry breaking[18–20] have been ubiquitously observed across $AV_3Sb_5$, $ScV_6Sn_6$, and FeGe, supporting the universal scenario of vHS-driven electronic ordering. On the other hand, $AV_3Sb_5$, $ScV_6Sn_6$, and FeGe possess quite distinct structural and magnetic properties: $AV_3Sb_5$ possesses a nonmagnetic V-kagome lattice in the unit cell, $ScV_6Sn_6$ possesses a nonmagnetic bilayer V-kagome lattice (see Fig. 1a,b), and FeGe possesses a magnetic Fe-kagome lattice that develops A-type antiferromagnetic ordering below $T_N$ = 400 K (Fig. 1f-h). Moreover, the charge order in $ScV_6Sn_6$ sets in with $\sqrt{3} \times \sqrt{3} \times 3$ periodicity (Fig. 1a), distinct from the $2 \times 2 \times 2$ charge orders found in $AV_3Sb_5$ and FeGe (Fig. 1f). These diverging properties suggest that more material-specific mechanisms, such as steric effects from small size of Sc in $ScV_6Sn_6$ and spin-phonon coupling in FeGe, might be key factors driving charge order in each compound.[21,22] Nevertheless, despite wide efforts, the exact nature of charge order in kagome lattice materials as well as the respective roles of electronic, structural, and magnetic degrees of freedom remain elusive, calling for a new experimental approach.

In general, the nature of complex emergent orders in quantum materials is hard to discern in static experiments, where all contributing degrees of freedom lie in equilibrium. One promising approach to tackle this challenge is using time-resolved techniques, which drive the system far-out-of-equilibrium using ultrafast photoexcitation, decouple different degrees of freedom, and monitor subsequent dynamics.[23] In this dynamical process, the interplay between different degrees of freedom and order parameters is directly imprinted in the time traces, making time-resolved techniques a powerful tool to unravel the nature and interplay of complex emergent orders. Among various time-resolved techniques, femtosecond X-ray scattering at an X-ray free electron laser (XFEL) has a special merit, as it grants both energy- and momentum-resolution, allowing direct access to the dynamics of finite wave vector order parameters, such as charge and spin orders in kagome lattice materials. However, despite its promise, time-resolved X-ray scattering investigations on kagome materials have been limited thus far.[24]

In this work, employing time-resolved X-ray diffraction (Tr-XRD) at XFEL, we demonstrate that the nature of charge orders in kagome systems, highly controversial in static experiments, can be directly determined in the time-domain. Specifically, by comparing nonmagnetic and magnetic kagome systems, $ScV_6Sn_6$ and FeGe (Fig. 1), we discovered that the dynamics of charge order peaks are fundamentally different (Fig. 2), directly revealing distinct underlying mechanisms and free energy landscapes dictating charge ordering in each compound (Fig. 3). In particular, while the dynamics in $ScV_6Sn_6$ are characterized by ultrafast melting and coherent amplitudon oscillations typical of phonon-coupled charge-ordered systems, the dynamics in FeGe display a resilient metastable behavior hitherto unobserved, which we interpret as a signature of unconventional, magnetism-interlocked nature of charge order in this kagome magnet. Our results not only uncovered rich mechanisms behind charge order formation in different kagome lattice materials, but also highlight femtosecond X-ray scattering as a powerful tool to decipher these underlying physics.

We start by discussing the static charge order peaks in $ScV_6Sn_6$ and FeGe measured without pump pulses at XFEL, as shown in Fig. 1d,i (see also Supplementary Section 1 for detailed experimental geometry). Clear superlattice reflections were observed at $Q = (1/3, 1/3, 2/3)$ in ScV$_6$Sn$_6$ and at $Q = (1/2, 0, 1/2)$ in FeGe, consistent with the $\sqrt{3} \times \sqrt{3} \times 3$ and $2 \times 2 \times 2$ charge orders in each compound as established in previous studies.[9,10] These superlattice peaks disappear upon increasing the temperature beyond $T_{CO}$ (97 K in ScV$_6$Sn$_6$ and 110 K in FeGe), confirming that the peaks originate from charge ordering rather than external artifacts such as higher harmonics in XFEL pulses. In both ScV$_6$Sn$_6$ and FeGe, the charge order distortion is substantial (with charge order intensity reaching 1/20 of the Bragg peak), the correlation is long-ranged ($\xi_{CO} \approx 102$ nm and 79 nm, respectively), and the temperature dependence follows a weak first order-like transition at $T_{CO}$ (Fig. 1e,j). These similarities in charge order behavior in static diffraction underscore the challenge in resolving their potentially distinct natures under equilibrium conditions.

In contrast to these similar static behaviors, the charge order peaks in ScV$_6$Sn$_6$ and FeGe display markedly different dynamics upon ultrafast photoexcitation with a 1.55 eV optical pulse, as shown in Fig. 2. First, in ScV$_6$Sn$_6$, the charge order peak undergoes ultrafast melting within 160 fs, followed by the launch of a 1.48 THz coherent oscillation that persists for over 5 ps (Fig. 2a). Such ultrafast melting and coherent mode oscillations are archetypal dynamics of phonon-coupled charge-ordered materials. We fit the time traces using a standard function,

$$I = 1 - \frac{1}{2}\left[1 + \text{erf}\left(\frac{t-t_0}{t_m/2\sqrt{\ln 2}}\right)\right] \times \left[A\,e^{-\frac{t-t_0}{t_r}} + B\,e^{-\frac{t-t_0}{t_d}}\cos(2\pi f t + \phi) + C\right],$$

from which we extracted relevant parameters as summarized in Fig. 2b-e and Supplementary Fig. S2. With increasing the pump fluences $fl$, the melting of charge order $A$ starts to saturate above $fl_c = 0.44$ mJ/cm$^2$ (Fig. 2c), indicating that a nonthermal phase transition from the normal to the charge-ordered state has been achieved. Accordingly, the melting time $t_m$ exhibits critical slowing down at $fl_c$, consistent with the enhanced fluctuation at the phase transition (Fig. 2b).[25] Meanwhile, the coherent oscillation displays a strong softening of frequency $f$ and enhanced damping $1/t_d$ with increasing the temperature toward $T_{CO}$ (see the inset in Fig. 2a), suggesting it as the amplitude mode of charge order in ScV$_6$Sn$_6$. The cosine-like phase of oscillation ($\phi \approx 0$, Fig. 2e) and the melting time ($t_m \approx 160$ fs) about one-quarter of the oscillation period ($1/f \approx 670$ fs) further support this assignment of the 1.48 THz oscillation to the amplitude mode, coherently excited via the

'displacive excitation of coherent phonon (DECP)' mechanism.[26] We note that this amplitude mode of charge order, also observed in recent time-resolved reflectivity measurements on ScV$_6$Sn$_6$,[27] has not been detected in the related nonmagnetic kagome system $A$V$_3$Sb$_5$.[24] After full melting of the charge order phase ($fl \geq fl_c$), we further observed a new, strongly damped oscillation corresponding to the oscillation with respect to the new equilibrium atomic positions in the normal state (see bottom panel of Fig. 2a). Overall, the order parameter dynamics observed in ScV$_6$Sn$_6$ can be considered an archetype of dynamics in phonon-coupled charge-ordered materials, resembling those established in other charge order systems, including Cr,[28] $RE$Te$_3$ ($RE$: rare earth),[29] VSe$_2$,[30] and K$_{0.3}$MoO$_3$,[31] to name a few.

As opposed to this standard behavior in ScV$_6$Sn$_6$, we observed markedly different dynamics in the magnetic kagome metal FeGe. As shown in Fig. 2f,g, the dynamics in FeGe is characterized by a surprising absence of ultrafast response within $t <$ 5 ps temporal window, while only showing a gradual decrease in charge order intensity over a much longer $\approx$ 30 ps timescale. Such absence of an ultrafast response in FeGe is thoroughly confirmed under various experimental conditions, including different pump polarizations, pump fluences, temperatures, and peak indices as shown in Fig. 2f,g and Supplementary Fig. S3. Considering that the electronic temperature of FeGe transiently increases to above 1100 K within a few hundreds of fs after the arrival of pump pulses (see time-resolved reflectivity measurements in Fig. 2f and two temperature model calculation in Supplementary Fig. S5), this absence of an ultrafast response indicates a surprising stability of charge order in FeGe against a nonthermal increase in electronic temperature, in stark contrast to the ultrafast melting triggered in ScV$_6$Sn$_6$. Given that the charge order in FeGe has a relatively low equilibrium $T_{CO}$ = 110 K, this behavior also implies different thermal and nonthermal pathways in FeGe. Up to our knowledge, such resilience to photoexcitation has not been reported in any other charge order systems: all charge-ordered materials studied so far using either Tr-XRD or ultrafast electron diffraction – including transition metal dichalcogenides, rare earth tri- and tetra-tellurides, correlated oxides, one-dimensional systems, and elemental metals – all exhibited ultrafast melting on subpicosecond timescales, as we catalogued in Supplementary Table S2. As elaborated below, this unique dynamics of FeGe is a direct consequence of the unconventional nature of charge order in this kagome magnet, inherently interlocked with the background magnetism.

The distinct charge order dynamics in ScV$_6$Sn$_6$ and FeGe can be further highlighted by comparing the temporal evolution of peak profiles (Fig. 2h,i). In ScV$_6$Sn$_6$, the charge order peak displays an ultrafast reduction of overall intensity within a subpicosecond timescale without any noticeable change in peak position (Fig. 2h). In contrast, in FeGe, the main response of charge order peak to photoexcitation is a shift of the peak center to lower wave vector over tens of ps, without ultrafast response on subpicosecond timescales (Fig. 2i). The latter behavior indicates that the gradual decrease in diffraction intensity observed in FeGe is predominantly driven by thermal lattice expansion and associated change in the Debye-Waller factor,[32] while an electronic response is absent. Lastly, we did not observe noticeable pump-induced change in peak width in both ScV$_6$Sn$_6$ and FeGe, indicating the absence of dramatic change in the underlying domain structure or topological defects under our experimental conditions.

Having fully characterized the charge order dynamics, we now demonstrate that the markedly different time-domain behavior of ScV$_6$Sn$_6$ and FeGe directly reflect the distinct nature of charge orders in these two kagome compounds. To illustrate this, we first compare the free energy landscapes underlying charge order transitions in ScV$_6$Sn$_6$ and FeGe, as shown in Fig. 3a-d. In ScV$_6$Sn$_6$, phonon dispersion calculations in Fig. 3a reveal a negative-frequency phonon mode, indicating that the pristine structure of ScV$_6$Sn$_6$ is unstable toward developing a finite distortion.[33] Accordingly, the free energy profile of ScV$_6$Sn$_6$ exhibits a *double-well potential* (Fig. 3b), where charge order phases at finite distortions are stabilized due to the instability of the pristine phase. We note that the calculated potential profile of ScV$_6$Sn$_6$ is asymmetric, which is likely the origin of weak first-order-like transition observed at $T_{CO}$ (Fig. 1e). In contrast, in FeGe, no negative-frequency phonon mode is observed in the phonon dispersion calculations (Fig. 3c), indicating that the pristine structure is dynamically stable if only the elastic energy is considered (orange dashed line in Fig. 3d). However, FeGe is a magnetic system, and it turns out that the magnetic exchange energy is largely saved with increasing distortion from the pristine phase (blue dashed line in Fig. 3d). Microscopically, this magnetic energy saving originates from the weakening of Fe-Ge1 orbital hybridization with increasing distortion, which enhances local correlation and magnetic moments on Fe atoms.[22,34,35] As a result of this delicate interplay between elastic and magnetic energy, FeGe exhibits a characteristic *triple-well potential* landscape, distinct from the double-well potential found in ScV$_6$Sn$_6$ and other charge-ordered materials. The physical implication of this triple-well potential is that, in addition to the already stable pristine structure at zero distortion, there appears a competing stable structure at finite distortion, whose stability is underpinned by the background magnetic order. In essence, this mechanism indicates that the charge order in FeGe possesses highly nontrivial origin, arising from the interplay of lattice distortion, electronic correlation, and magnetism, instead of the instability in the pristine phase.

We note that this magnetic energy saving mechanism has been theoretically proposed as one of the candidate mechanisms for charge order in FeGe,[22,34,35] along with other competing scenarios.[12,36,37] However, experimental tests of this proposal have been challenging, largely because static experiments conducted so far only characterized the state at the minimum of the free energy, without assessing the full potential landscape. Unlike static measurements, the dynamics probed in our Tr-XRD are directly sensitive to the underlying potential configurations. For instance, with increasing electronic temperature by photoexcitation, the double-well potential in $ScV_6Sn_6$ gets strongly renormalized, with the position of potential minimum shifting to smaller distortion (see red-solid lines in Fig. 3e). This shift of the potential minimum is a prerequisite for the DECP mechanism,[26] driving the ultrafast decay of the order parameter and launching coherent amplitudon oscillations, as schematically illustrated in Fig. 3e. The estimation of the amplitudon frequency based on the curvature of the first-principle potential profile yields $f_{DFT}$ = 1.50 THz, in close agreement with the experimental $f$. In contrast, for the triple-well potential in FeGe, our calculations reveal that the position of the local minimum of the free energy remains surprisingly unaltered with increasing electronic temperature, as marked with red-solid line in Fig. 3f. Since the order parameter keeps staying at the minimum of the free energy even after photoexcitation, there is no "force" or "impetus" to drive a change or oscillation in the order parameter. This fully explains the resilient charge order behavior observed in FeGe as a direct consequence of the underlying triple-well potential landscape, arising from the novel magnetism-stabilized charge order mechanism in this material.

For a more quantitative description of the above physics, we also constructed the Euler-Lagrange equation for the order parameter $\psi$:

$$\frac{d\psi^2}{d^2t} = -2\frac{1}{\gamma}\frac{d\psi}{dt} - \frac{1}{m_{eff}}\frac{dF(\psi, T_e(t))}{d\psi},$$

using DFT calculated free energy functionals $F(\psi, T_e(t))$ in Fig. 3e,f as an input. Here, $\gamma$ is the phenomenological damping constant, $m_{eff}$ is the effective mass associated with $\psi$, and $T_e(t)$ is the time-dependent electron temperature, accounting for the pump pulse at time zero (see Methods and Supplementary Section 5 for details). As shown in Fig. 3g,h, the simulated $\psi(t)$ features ultrafast melting and coherent amplitude mode oscillation in ScV$_6$Sn$_6$ and resilience to photoexcitation in FeGe, closely reproducing the experimental dynamics. This confirms that the distinct dynamics observed in our Tr-XRD (Fig. 2) essentially probes the different underlying potential landscapes of ScV$_6$Sn$_6$ and FeGe (Fig. 3). We reiterate that these different potential landscapes are direct consequences of the fundamentally different driving mechanisms of charge orders in ScV$_6$Sn$_6$ and FeGe, where structural and magnetic degrees of freedom play a key role, respectively. Our study thus provides an important model case where the distinct nature and driving mechanisms of electronic orderings, challenging to distinguish in equilibrium, are directly decoded in the time-domain through dramatically different dynamics.

We now discuss the implications of our study in terms of the broader kagome metal physics. In the canonical charge-ordered kagome system, $A$V$_3$Sb$_5$, a consensus has been developed that the charge order has an electronic origin, arising from instabilities associated with van Hove singularities at the Fermi level.[38] Depending on tuning parameters, this van Hove singularity instability can also stabilize other competing symmetry broken phases alongside the charge order, such as superconductivity and electronic nematicity. Accordingly, these competing electronic phases are experimentally observed in the phase diagram of $A$V$_3$Sb$_5$.[8,39] Meanwhile, ScV$_6$Sn$_6$ possesses the same nonmagnetic V-based kagome lattice and similar electronic structure with $A$V$_3$Sb$_5$, yet our observation of the unstable phonon mode and strong amplitude mode oscillation suggests the dominant role of structural rather than electronic instability. This may explain the dichotomy between the $A$V$_3$Sb$_5$ and ScV$_6$Sn$_6$ phase diagrams, where competing electronic phases are absent in the latter (Fig. 1c).[17,40] Finally, the unique dynamics observed in our Tr-XRD demonstrate that the charge order in magnetic kagome system FeGe has a distinguished origin, underpinned by the background antiferromagnetic order. In this context, charge order in the three kagome lattice materials, $A$V$_3$Sb$_5$, ScV$_6$Sn$_6$, and FeGe, have all different origins, with electronic, structural, and magnetic degrees of freedom playing key roles, respectively. It is truly remarkable that an ostensibly similar phenomena (charge order in kagome lattice geometry) in fact originate from such diverse underlying physics. We propose that applying a similar time-resolved approach to other charge-, stripe-, or spin stripe-ordered kagome systems – including 4d kagome metals LaRu$_3$Si$_2$ and LuNb$_6$Sn$_6$, and the strongly correlated kagome system CsCr$_3$Sb$_5$ – would provide further insights into the diverse nature of electronic orderings found within the wide family of kagome lattice materials.[41–43]

We conclude our study by highlighting different thermal and nonthermal pathways in kagome metals, taking FeGe as a prominent example (see schematics in Fig. 4a,b). In the equilibrium/thermal case, the system always resides in the *global* minimum of the free energy. This indicates that with increasing temperature, the transition to the non-distorted normal state occurs as soon as $F(\psi = 0) < F(\psi = 1)$, which defines the relatively low charge order transition temperature $T_{CO}$ = 110 K in FeGe (Fig. 4a). In contrast, in the out-of-equilibrium/nonthermal regime explored in our ultrafast experiments, the charge order state remains *metastable* up to much higher (electronic) temperature $T_e$ = 4000 K, far after the global minimum has shifted to the $\psi = 0$ (Fig. 4b). This is due to the potential barrier $W$ between the normal and charge-ordered states, preventing thermal fluctuations within the measurement timescale. In this context, the resilient charge order in FeGe observed in our ultrafast experiments is likely the metastable state trapped in the local minimum instead of the global minimum of the free energy. This unique setting in FeGe also provides a promising platform to drive and engineer various light-induced out-of-equilibrium phenomena. For example, one can imagine starting in the normal state ($T > T_{CO}$, $\psi = 0$) and driving the order parameter using, for example, intense THz pulses and nonlinear phononics,[44] to realize a metastable charge order state ($T > T_{CO}$, $\psi = 1$) at temperatures far above the equilibrium $T_{CO}$. Alternatively, one can also imagine more dramatically altering the free energy landscape of FeGe by resonantly melting its magnetism – considering that magnetic ordering is a prerequisite for stabilizing the charge order in FeGe, this will offer a novel pathway to control charge order by modulating spin degrees of freedom in kagome magnets. Altogether, our results invite for future time-resolved studies in kagome lattice materials not only to decode the unconventional nature of electronic orders in time-domain, but also to engineer and drive novel out-of-equilibrium phenomena in this promising lattice platform.

**Methods**

**ScV$_6$Sn$_6$ sample growth and characterizations** Single crystals of ScV$_6$Sn$_6$ were grown by typical self-flux methods. Scandium pieces (99.9 % Research Chemicals), Vanadium pieces (99.7 % Alfa Aeser), and Sn ingot (99.99 % Alfa Aeser) were put in the Alumina crucible with frit disc, then sealed in Ar-gas purged evacuated quartz tube. Ampule was heated at 1100 °C for 24 hrs, then slow cooled to 800 °C with 1~2°C/hr cooling ratio. To remove the flux, ampule was centrifuged at 800 °C. Elemental ratios were confirmed using energy dispersive spectroscopy. Electrical Resistivity measurements was performed with Physical Properties Measurement System (PPMS, Quantum design) using a conventional 4 probe method, confirming expected charge order transition at $T_{CO}$ = 97 K.

**FeGe sample growth and characterizations** B35-type FeGe single crystals were grown by the chemical vapor transport (CVT) method using iodine as the transport agent. The stoichiometric mixture of iron powders (99.99 %) and germanium powders (99.999 %) was sealed in an evacuated quartz tube with additional iodine as transport agents. The quartz tube was annealed in a horizontal two-zone furnace with a temperature gradient 600 °C (source) to 550 °C (sink) for three weeks to grow single crystals. After being cooled naturally to room temperature, FeGe crystals with a typical dimension of 2×2×2 mm$^3$ can be obtained in the middle of the quartz tube. The postgrowth annealing treatments were adopted to enhance the charge order volume fraction. The as-grown crystals were sealed into a quartz tube under high vacuum and put into a box furnace. After being held at 320 °C for a week, the quartz tube was quickly taken out of the furnace and quenched in room-temperature water. The annealed crystals show long-ranged charge order with $T_{CO}$ = 110 K, as confirmed by our homelab and synchrotron XRD measurements.

**Static XRD measurements at synchrotron** Prior to XFEL experiments, static XRD experiments were conducted at the hard X-ray diffraction beamline (3A) of Pohang Light Source (PLS-II) to characterize equilibrium charge orders in ScV$_6$Sn$_6$ and FeGe. Samples were cleaved in air right before loading into the vacuum chamber. We characterized charge orders in ScV$_6$Sn$_6$ and FeGe by measuring (1/3, 1/3, 19/3) and (1/2, 0, 3) charge order peaks, respectively. Temperature-dependent measurements were performed to confirm the first-order-like charge order transition at $T_{CO}$ = 97 K and 110 K as shown in Fig. 1e,j.

**Tr-XRD measurements at XFEL** Tr-XRD experiments on $ScV_6Sn_6$ and FeGe were conducted at the RSXS endstation of PAL-XFEL. 800 nm optical pulses with a pulse duration 100 fs and X-ray pulses with a duration 80 fs were used as pump and probe pulses, respectively. The total temporal resolution of the experiments was 130 fs. The repetition rate of the pump pulse was set to half (30 Hz) of that of the probe pulse (60 Hz), so that half of the XFEL pulses probed the photoexcited state while the remaining half measured the unperturbed state as a reference. The reference signals were used to monitor stability during the experiments, as well as to assess static heating and beam damage on the sample. The optical pump pulses were injected nearly parallel to the X-ray pulses. The footprint of the pump beam (~500 × 500 μm$^2$) was set to be much larger than the probe beamspot (~100 × 200 μm$^2$) to ensure quasi-uniform photoexcitation across the entire probe volume. We used the effective fluence values throughout the Main text corrected by considering the angle of laser incidence. The maximum effective pump fluence used in our experiments was 2 mJ/cm$^2$, above which we observed pump-induced irreversible damage on both $ScV_6Sn_6$ and FeGe. The diffraction experiments were carried out in four-circle geometry using avalanche photodiodes to detect diffracted photons with enhanced sensitivity. The polarization of the X-ray pulses was fixed parallel to the scattering plane ($\pi$ polarization). For $ScV_6Sn_6$, we used 900 eV photons to measure (1/3, 1/3, 2/3) charge order peak, while for FeGe, 1100 eV and 1720 eV photons were used to measure the (1/2, 0, 1/2), (1/2, 0, 1), and (0, 0, 1/2) charge order peaks. We obtained qualitatively similar dynamics within each sample, regardless of the peak index (Supplementary Fig. S2). The samples were cooled using an open-cycle LHe cryostat within an ultrahigh vacuum chamber with a base pressure around 10$^{-9}$ Torr. The results presented in the Main text were obtained at the base temperature, 30 K, unless specified. Two $ScV_6Sn_6$ crystal and three FeGe crystals were measured over two different beamtimes, yielding consistent results.

**Time-resolved reflectivity measurements**
Time-resolved Reflectivity (Tr-R) experiments were performed at the T-ReX laboratory (Elettra, Trieste) using a Ti:sapphire amplified laser system (Coherent RegA), delivering ≈ 50 fs light pulses at a wavelength of 800 nm (1.55 eV), at a repetition rate of 250 kHz. A fraction of the laser output was used as a pump. The probe wavelengths were derived from a broadband supercontinuum pulse. Upon reflection from the sample, the broadband probe beam was split in two beams, which were filtered with interferential filters at 650 nm and 1100 nm (Thorlabs, 10 nm bandwidth) and detected with a Si and an InGaAs photodiode respectively. The ultrafast dynamics at the two wavelengths were acquired simultaneously using two lock-in amplifiers. Experiments were performed in a UHV open-cycle cryostat. The representative Tr-R data measured with 1100 nm probe is shown in Fig. 2f for comparison with Tr-XRD data. The full temperature- and fluence-dependent Tr-R datasets are discussed in Supplementary Section 4 and Supplementary Fig. S6.

**DFT calculation** DFT calculations were performed using the Vienna ab initio simulation package.[45,46] The generalized-gradient approximation Perdew-Burke-Ernzerhof exchange-correlation functional was chosen to calculate the exchange-correlation energy.[47] The pseudopotential was defined based on the projector augmented-wave method.[48] For $ScV_6Sn_6$, we used the experimentally refined $\sqrt{3} \times \sqrt{3} \times 3$ charge order superstructure in Ref.[10] as inputs for DFT calculation. For the free energy landscape calculation, we artificially modulated the degree of distortion from –175 % to +175 % of the equilibrium structure in 12.5 % step. Total energy of each structure is calculated using a Γ-centered $4 \times 4 \times 1$ $k$-point mesh with kinetic energy cutoff at 400 eV. For FeGe, we adopted the $2 \times 2 \times 2$ charge order superstructure in Ref.[9], and used a Γ-centered $4 \times 4 \times 4$ $k$-point mesh with kinetic energy cutoff at 450 eV for the total energy calculation. On-site Hubbard interaction $U = 1$ eV was applied for FeGe calculation to account for the magnetic ground state. To simulate the effect of pump fluences, we repeated free energy calculation with various smearing factors at the Fermi-Dirac distribution (0.05, 0.10, 0.15, 0.2, 0.25, 0.3, 0.35, and 0.375 eV), which mimics the effect of enhanced electronic temperature.

**DFPT calculation** Phonon dispersions in Fig. 3a,c were calculated using density functional perturbation theory (DFPT) functional in VASP. Input parameters for $3 \times 3 \times 2$ supercell of $ScV_6Sn_6$ and $2 \times 2 \times 2$ supercell of FeGe were generated using Phonopy software.[49,50]

**Simulation of the order parameter dynamics using first-principle free energies** We simulate the order parameter dynamics of $ScV_6Sn_6$ and FeGe using the Euler-Lagrange equation of motion for the order parameter $\psi$:

$$\frac{d\psi^2}{d^2t} = -2\frac{1}{\gamma}\frac{d\psi}{dt} - \frac{1}{m_{eff}}\frac{dF(\psi, T_e(t))}{d\psi}.$$

Here $\gamma$ is the phenomenological damping constant, $m_{eff}$ is the effective mass of the system, $T_e(t)$ is a time-dependent electron temperature accounting for the effect of pump pulse at time zero, and $F(\psi, T_e)$ is a free energy functional. We used a weight-averaged mean square distortion and total mass in the unit cell,

$$\psi = \sqrt{\frac{\sum m_i \psi_i^2}{\sum m_i}}, \qquad m_{eff} = \sum m_i,$$

as the order parameter and effective mass for the Euler-Lagrange equation. The free energy profile $F(\psi, T_e)$ is obtained using DFT. Note that DFT calculation only provides $F(\psi, T_e)$ for discrete values of $\psi$ and $T_e$. We thus fit $F(\psi, T_e)$ with a generic bivariate polynomial and used the obtained fitting function $F^{fit}(\psi, T_e)$ for the analytic simulation (see Supplementary Section 5 for details). The effect of pump-pulse at time-zero is considered as time-dependent electron temperature

$$T_e(t) = T_0 + \frac{1}{2} fl \left[ 1 + \mathrm{erf}\left(\frac{t}{\sqrt{2} t_{rise}}\right) \right] \left( e^{-\frac{t}{t_{decay}}} + C' \right),$$

where $T_0$ is the base temperature, $fl$ is the pump fluence, $t_{rise}$ and $t_{decay}$ represent the rise and decay timescales of electron temperature, and $C'$ is the thermal offset after the pump. With these inputs, the Euler-Lagrange equation was numerically solved using a standard differential equations solver. The order parameter values at the equilibrium, defined by $\frac{dF^{fit}(\psi, T_0)}{d\psi} = 0$, were used as the initial condition. We refer to the Supplementary Section 5 for further details of the simulation.

**Acknowledgements**

We acknowledge Pohang Accelerator Laboratory for provision of synchrotron and XFEL beamtime. The Tr-XRD experiments were performed at the RSXS endstation (proposal number: 2023-2nd-SSS-015, 2024-2nd-SSS-013) of the PAL-XFEL, funded by the Korea government (MSIT). S.L., B.L., C.W., H.J., and J.-H.P. were supported by National Research Foundation (NRF) of Korea grant No. RS-2022-NR068223. X.W and A.W were supported by the Natural Science Foundation of China (Nos. 12474142) and the New Chongqing Youth Innovative Talent Project (No. 2024NSCQ-qncxX0474). K.S. acknowledges the support from the Air Force Office of Scientific Research (Grants No. FA9550-21-1-0168 and No. FA9550-23-1-0161). J.K. acknowledges a PIURI postdoctoral fellowship from POSTECH (grant No. 2021R1A6A1A10042944). M.K. acknowledges a Kavli postdoctoral fellowship from Kavli Nanoscience Institute at Cornell.

**Author contributions**

J.-H.P. and M.K. conceived the project; S.L., B.L., and J.K. conducted the static and time-resolved XRD experiments and analyzed the resulting data with help from H.J., G.K., and S.-Y.P.; F.C. performed time-resolved reflectivity measurements; S.L. performed the theoretical calculations; C.W. synthesized and characterized the $ScV_6Sn_6$ crystals.; X.W. and A.W. synthesized and characterized the FeGe crystals.; S.L. and M.K. wrote the manuscript with input from all coauthors.

**Data availability**

The datasets presented within this study are available from the corresponding authors upon reasonable request.

**Competing interests**

The authors declare no competing interests

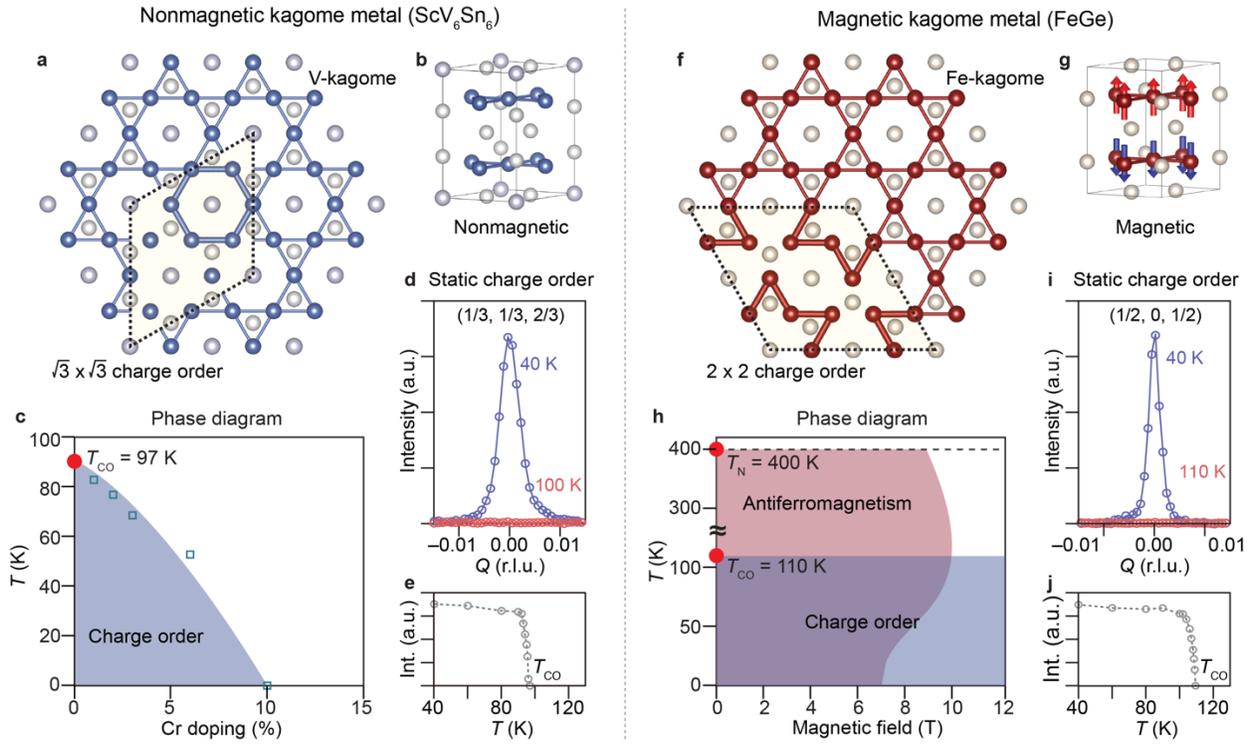

**Figure 1 | Charge orders in nonmagnetic and magnetic kagome systems. a,b,** In-plane kagome lattice structure and out-of-plane stacking structure of nonmagnetic kagome system $ScV_6Sn_6$, respectively. The overlaid parallelogram in a represent $\sqrt{3} \times \sqrt{3}$ charge order distortions. **c,** Schematic phase diagram of $ScV_6Sn_6$ featuring charge order phase below $T_{CO}$ = 97 K. **d,e,** (1/3, 1/3, 2/3) charge order peak profile of $ScV_6Sn_6$ and its temperature evolution across $T_{CO}$. **f,g,** In-plane kagome lattice structure and out-of-plane stacking structure of magnetic kagome system FeGe, respectively. The overlaid parallelogram in a represent $2 \times 2$ charge order distortions. **h,** Schematic phase diagram of FeGe. In contrast to $ScV_6Sn_6$, in FeGe, charge order ($T_{CO}$ = 110 K) develops within the background antiferromagnetic order ($T_N$ = 400 K). **i,k,** (1/2, 0, 1/2) charge order peak profile of FeGe and its temperature evolution across $T_{CO}$.

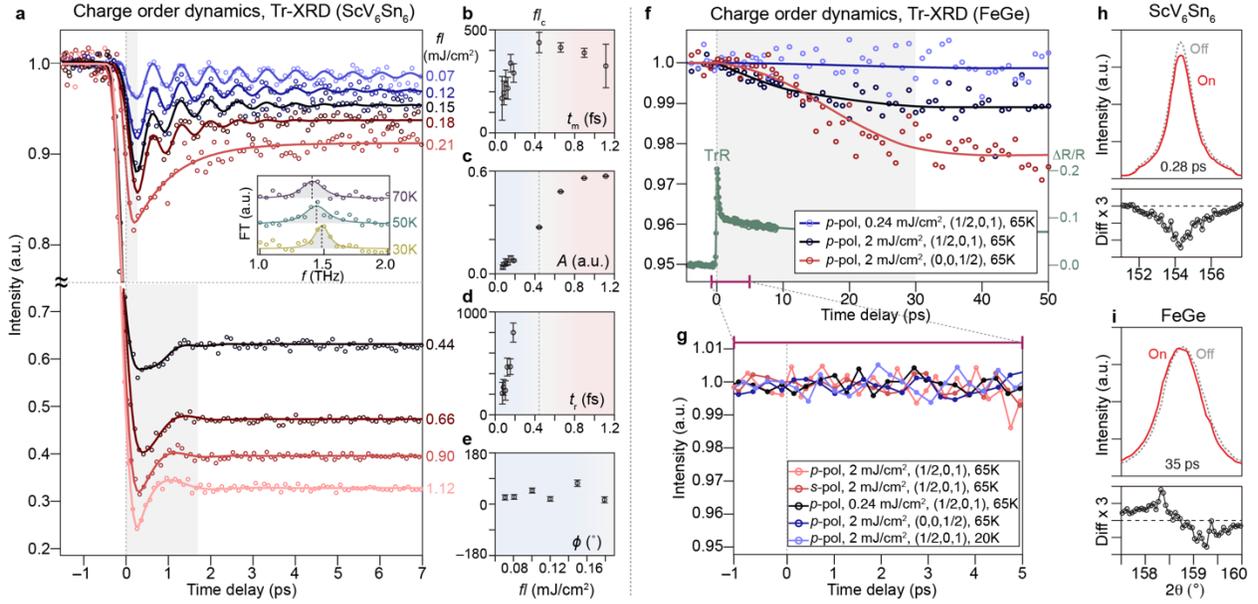

**Figure 2 | Distinct charge order dynamics in nonmagnetic kagome metal ScV$_6$Sn$_6$ and magnetic kagome metal FeGe. a,** Dynamics of the charge order peak in ScV$_6$Sn$_6$ as a function of pump fluence $fl$ (mJ/cm$^2$). Overlaid solid lines are fit to a standard function described in the main text. The inset shows the Fourier transform of the oscillatory component of the signal as a function of temperature. **b-e,** Melting time $t_m$, melting amplitude $A$, recovery time $t_r$, and phase of coherent oscillation $\phi$ obtained from the fit described in the Main text. **f,** Dynamics of charge order peaks in FeGe measured under various experimental conditions. Solid lines are guides to the eye. Also shown with a green-dotted line is the complementary time-resolved reflectivity data on FeGe, showing ultrafast response of the electronic bath. **g,** A zoomed-in view of the short-delay region in f, highlighting the absence of an ultrafast response in FeGe independent of experimental conditions. **h,** Pump-induced change in the (1/3, 1/3, 2/3) charge order peak profile of ScV$_6$Sn$_6$ measured at a 0.28 ps time delay with pump fluence $fl$ = 0.12 mJ/cm$^2$. **i,** Pump-induced change in the (1/2, 0, 1) charge order peak profile of FeGe measured at a 50 ps time delay with pump fluence $fl$ = 2 mJ/cm$^2$. The bottom panels in a,b show the difference between pumped (on) and unpumped (off) signal.

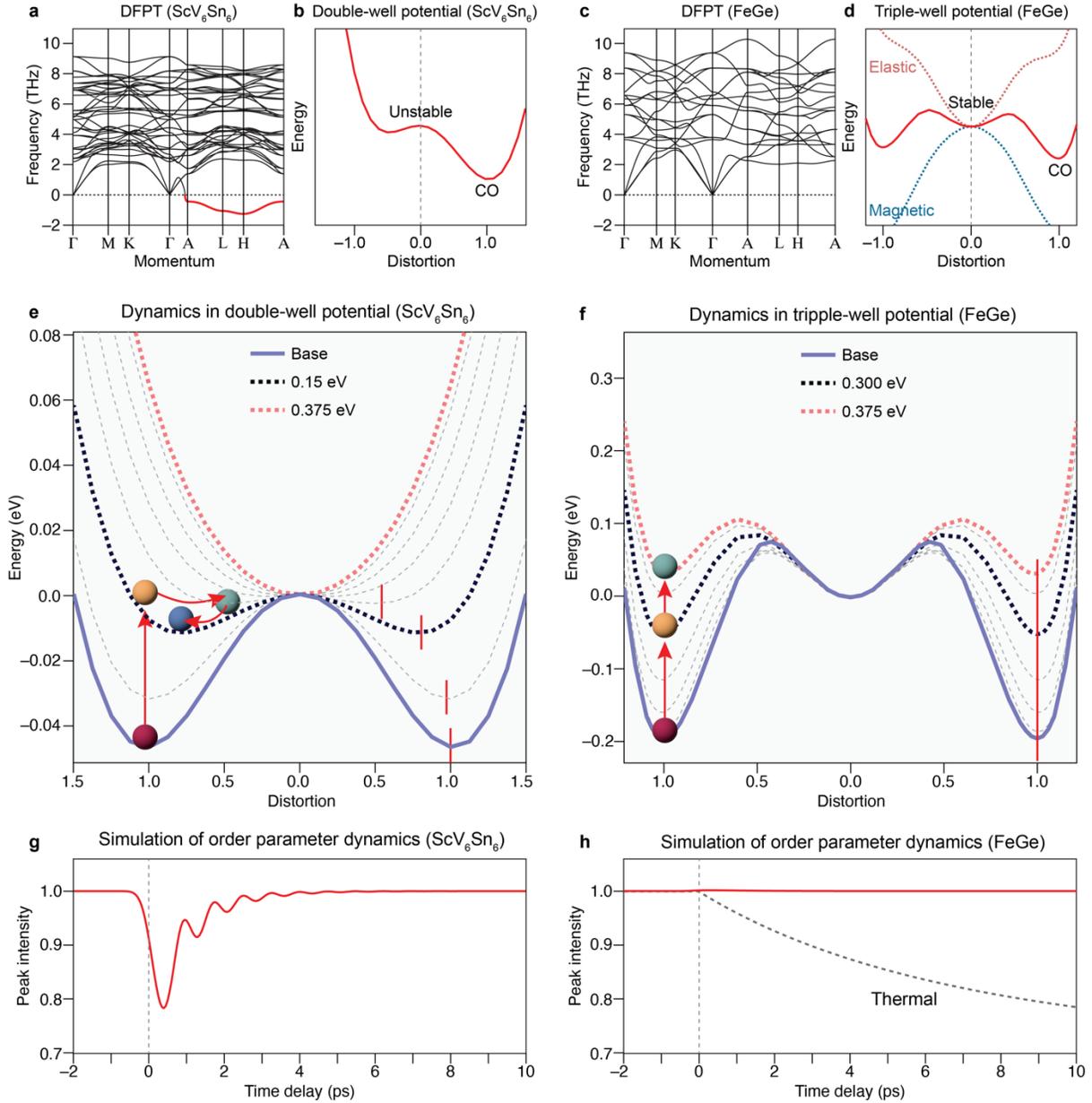

**Figure 3 | Distinct free energy landscapes and order parameter dynamics in ScV$_6$Sn$_6$ and FeGe. a,** Phonon dispersion of ScV$_6$Sn$_6$ calculated using density functional perturbation theory. The negative-energy phonon branch, highlighted in red, indicates structural instability. **b,** The resultant double-well potential profile of ScV$_6$Sn$_6$. **c,d,** The corresponding phonon dispersion and potential profile of FeGe. Despite the absence of phonon instability (c), the charge order state at finite distortion develops through a magnetic exchange energy saving mechanism, resulting in a triple-well potential profile in FeGe. **e,f,** Evolution of the potential profile as a function of electron temperature in ScV$_6$Sn$_6$ and FeGe, respectively. For simplicity, we only discuss the potential profiles at positive distortions (symmetrized) in e,f, while providing full potential profiles in Supplementary Fig. S9. Vertical red lines in e,f mark the minimum positions of the potential wells. Overlaid schematics depict the order parameter dynamics upon ultrafast photoexcitation. **g,h,** Simulations of order parameter dynamics in ScV$_6$Sn$_6$ and FeGe, respectively. The black dashed line in h represents the effect of lattice thermalization, capturing the slow decay of peak intensity observed in FeGe.

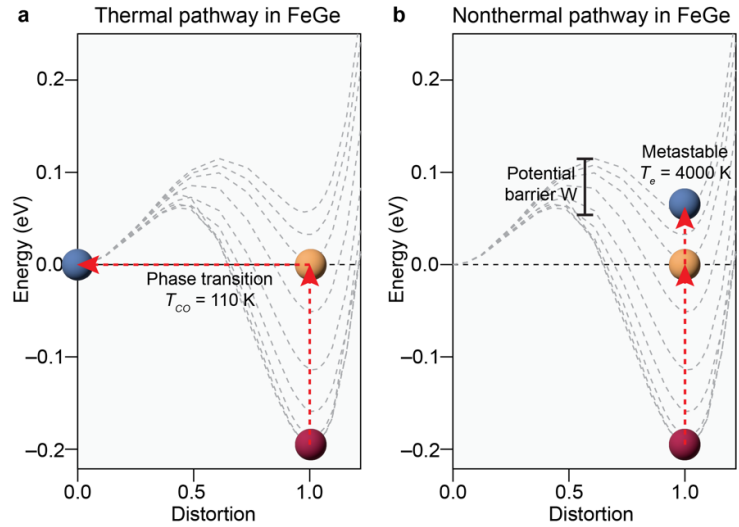

**Figure 4 | Thermal and nonthermal pathways in FeGe. a,b,** Schematic illustrations of thermal and nonthermal pathways in FeGe accessed in static and time-resolved experiments, respectively. The grey dashed curves in the background represent the electron temperature-dependent potential profile of FeGe, adapted from Fig. 3f. Overlaid schematics depict the evolution of order parameter through thermal and nonthermal pathways, respectively.